\documentclass[copyright,creativecommons]{eptcs}
\providecommand{\event}{HSB 2012} 
\usepackage{breakurl}             
\pdfoutput=1
\usepackage[utf8]{inputenc}
\usepackage{microtype}
\usepackage{enumerate,ifthen,xspace}
\usepackage{graphicx,color}
\usepackage{amsmath,amsfonts,amstext,amssymb}
\makeatletter\@ifclassloaded{llncs}{}{\usepackage{amsthm}}\makeatother
\usepackage{paralist}
\usepackage{aliascnt}
\usepackage{hyperref}
\usepackage[numbers]{natbib}
\usepackage{cleveref}
\usepackage{tikz}
\newenvironment{proofsketch}{\trivlist\item[]\emph{Proof Sketch}:}%
{\unskip\nobreak\hskip 1em plus 1fil\nobreak$\Box$
\parfillskip=0pt%
\endtrivlist}
\newcommand{\RH}{\ensuremath{RH}\xspace}
\newcommand{\HG}{\ensuremath{HG}\xspace}
\newcommand{\DHG}{\ensuremath{DHG}\xspace}
\newcommand{\CHAs}{CHAs\xspace}
\newcommand{\exec}{S}
\newcommand{\run}{r}

\ifx\pgfversion\undefined
  \input{pgfexternal.tex}
\else

  \usetikzlibrary{chains}
  \tikzstyle{hallmark}=[draw,rectangle]
  \tikzstyle{hallmarkchain}=[node distance=5mm,start chain,
  every node/.style={hallmark,on chain,join}, every join/.style={->}]
  \ifthenelse{\lengthtest{\pgfversion cm<1.18cm}}{
  }{
    \ifx\regeneratepgf\undefined\pgfrealjobname{\jobname}
    \else\pgfrealjobname{\regeneratepgf}
    \fi
  }
\fi

\newcommand{\dfn}[1]{\textbf{#1}}
\newcommand{\dfnless}[1]{\emph{#1}}
\newcommand{\mpunct}{\enspace}
\newcommand{\HH}{\ensuremath{\mathcal{H}}\xspace}

\newcommand{\AG}{\ensuremath{\mathsf{AG}}\xspace}

\newcommand{\Drugs}{\ensuremath{\mathcal{D}}\xspace}
\newcommand{\drugs}{\ensuremath{D}\xspace}
\newcommand{\drug}{\ensuremath{d}\xspace}
\newcommand{\cocktail}{\ensuremath{C}\xspace}

\newcommand{\Runs}{\ensuremath{\mathit{Runs}}\xspace}
\newcommand{\Runsf}{\ensuremath{\mathit{Runs}_{\mathsf{f}}}\xspace}

\newcommand{\Inv}{\ensuremath{\mathit{\ell}}\xspace}

\newcommand{\val}{\ensuremath{\mathsf{val}}\xspace}
\let\thera\pi
\let\candthera\Theta
\let\TT\theta

\makeatletter
\def\arrowtightfill@#1#2#3#4{%
$\m@th\thickmuskip0mu\medmuskip\thickmuskip\thinmuskip\thickmuskip
\relax#4#1%
\cleaders\hbox{$#4\mkern0mu#2\mkern0mu$}\hfill
#3$%
}
\def\dashrightarrowfill@{%
\arrowtightfill@\dabar@\dabar@
{\mathchar"0\hexnumber@\symAMSa 4B}%
}
\newcommand{\xdashrightarrow}[2][]{%
\ext@arrow 0359\dashrightarrowfill@{#1}{#2}%
}
\makeatother

\newcommand{\NAT}{\ensuremath{\mathbb{N}}\xspace}

\newcommand{\REAL}{\ensuremath{\mathbb{R}}\xspace}

\newcommand{\suchthat}{\,|\,}

\theoremstyle{theorem} 
\newtheorem{theorem}{Theorem}
\newtheorem{definition}{Definition}[theorem]
\newtheorem{example}[theorem]{Example}

\title{Towards Cancer Hybrid Automata}
\author{Loes Olde Loohuis
\institute{CUNY The Graduate Center\\ Department of Computer Science}
\email{l.oldeloohuis@gmail.com}
\and
Andreas Witzel \qquad\qquad Bud Mishra
\institute{NYU \\Courant Institute\\
}
\email{\quad awitzel@nyu.edu \quad\qquad mishra@nyu.edu}
}
\def\titlerunning{Cancer Hybrid Automata}
\def\authorrunning{L. Olde Loohuis, A. Witzel, B. Mishra}

\begin{document}

\maketitle

\begin{abstract}
  This paper introduces Cancer Hybrid Automata (\CHAs),
  a formalism to model the progression of cancers through discrete phenotypes.
  The classification of cancer progression using discrete states like stages and hallmarks has become common in the biology literature, 
  but primarily as an organizing principle, and not as an executable formalism.
  The precise computational model developed here aims to exploit this untapped potential, namely, through
  automatic verification of progression models (e.g., consistency, causal connections, etc.), classification of unreachable or unstable states
  and computer-generated (individualized or universal) therapy plans.
  The paper builds on a \emph{phenomenological} approach, and as such does not need to assume a model for the biochemistry of the underlying natural progression.
  Rather, it abstractly models transition timings between states as well as the effects of drugs and clinical tests,
  and thus allows formalization of temporal statements about the progression as well as notions of timed therapies.
  The model proposed here is ultimately based on \emph{hybrid automata},
  and we show how existing controller synthesis algorithms can be generalized to CHA models, so that therapies can be generated automatically. 
  Throughout this paper we use cancer hallmarks to represent the discrete states through which cancer progresses, 
  but other notions of discretely or continuously varying state formalisms could also be used to derive similar therapies. 
\end{abstract}

\section{Introduction}
\label{sec:introduction}

Cancer is generally thought of as a \emph{progressive disease} --
in particular, a disease which exhibits certain discernible cancer phenotypes (modeled as a finite set of \emph{discrete} states), through which it progresses towards a  terminal phenotype (e.g., metastasis).

Among other theories, this view is reflected in the so-called \emph{hallmarks of cancer} proposed by Hanahan and Weinberg \cite{hanahan_hallmarks_2000},
and it has become one of the predominant ways of thinking about cancer,
solidified through many further publications and experiments. 
A recent article by the same authors \cite{hanahan_hallmarks_2011}
reviews and consolidates the new insights of the last decade.
Similar models have also been explored by a mechanistic agent-based simulation in \cite{abbott2006simulating}.

According to the model proposed by Hanahan and Weinberg, tumors must necessarily acquire certain ``intermediate'' hallmarks
culminating in the ``final'' hallmarks of tissue invasion and metastasis.
As the authors write,
\begin{quote}
  Simply depicted, certain mutant genotypes confer selective advantage
  on subclones of cells, enabling their outgrowth and eventual
  dominance in a local tissue environment.  Accordingly, multistep
  tumor progression can be portrayed as a succession of clonal
  expansions, each of which is triggered by the chance acquisition of
  an enabling mutant genotype.
  \cite[p.~658]{hanahan_hallmarks_2011}
\end{quote}

The current list of cancer hallmarks includes the abilities to reproduce autonomously, to ignore anti-growth signals,
or to signal for formation of new blood vessels, as well as handful of other phenotypes.
Hallmarks can be obtained in various different orders, but not every order is viable.
Intuitively, a hallmark can be acquired by a dominant sub-population of cells
if it conveys a selective advantage compared to the other phenotypes acquired in that population.
For example, in a wildly growing cluster of cells, the ability to signal for new blood supply,
and thus nutrients, oxygen, and waste disposal, will allow the respective sub-population to outgrow the others.

Most hallmarks are acquired through mutations (point mutations, copy number changes or epigenetic modifications) of very specific sets of oncogenes and tumor suppressor genes. Thus, many of the targeted drugs, administered individually or combinatorially in a cocktail, which have been developed in recent years, aim to
influence the function of the products of these genes~\cite{luo_principles_2009} and thus cancer's evolution from specific hallmarks.
For example, the vascular endothelial growth factor (VEGF) signals for creation of new blood vessels (\emph{angiogenesis}),
and the drug Avastin inhibits the associated signaling pathway,
thus preventing growing tumors from obtaining the needed blood supply. While current therapies target only the observed hallmark at any instant, they rarely take into account the potential hallmarks that may evolve in the future and the temporal structure of the underlying evolution. By connecting therapy design to the theory of supervisory control of hybrid automata, we aim to build a framework for better therapy design (e.g., that avoids drug-resistance, exploits synthetic lethality, oncogene addiction, etc., and avoids undesirable side-effects on other organs). 

In this view of cancer, its progression through hallmarks and therapy bears a striking resemblance to
formal models of state-transition machines in computer science.

In this paper, we first present a logical framework called \emph{Cancer Hybrid Automaton} (CHA)
that allows us to formally capture cancer progression through accumulation of successive discrete states.
States in CHA models represent states of the progression, and directed edges among pairs of states define possible progression paths.
Drugs can then be thought of as inhibiting or prolonging specific transitions in the automaton. 
We then show how this approach enables us to formally describe cancer progression, automatically verify/model-check its temporal properties,
and manipulate its evolution to satisfy certain therapeutic goals. 
 
We illustrate our approach through a highly simplified running example of a cancer hybrid automaton in which states represent hallmarks, and progression paths 
represent successive hallmark acquisitions. However, the states of the automaton can represent any set of discrete states at varying levels of abstraction. Examples include stages of cancer, 
a set of affected pathways, 
and a set of specific genomic aberrations. 
By ignoring complex structures such as heterogeneity, geometry, circulating tumor cells, tumor growth dynamics, genomic instability at this point, we avoid obscuring the key ideas inherent to the therapy design algorithms. However, the framework is flexible enough to include such structures as well as detailed mechanistic models of the discrete states. 

\section{Overview}
\label{sec:overview}
The rest of this paper is organized as follows.
In \cref{sec:cancer-hallmark-automata}, we introduce a basic CHA formalism. 
In this section, a CHA is modeled as a \emph{finite non-deterministic automaton}. 
The edges, representing transitions from one progression state (e.g hallmarks) to the next, are labeled with drugs that can inhibit the transition. 
A \emph{therapy} is defined as a function that assigns a set of drugs to each finite progression history, or \emph{run}. 
An execution of a therapy is defined as a run of the CHA that respects the therapy, that is, no transition of the execution is inhibited by the therapy. 
Our model includes costs by associating a cost vector with each state and each cocktail.
Therapies may be selected by comparing costs of possible executions using a notion of Pareto dominance,
in addition to the required qualitative properties specified in CTL.

In \cref{sec:timed-cha} we extend the CHA framework to include real time. 
In this model, transitions take certain durations of time, and drugs can prolong (or stop) the transition process.
This is modelled using a hybrid automaton with multiple clocks \footnote{Hence the term \emph{hybrid} in `cancer hybrid automaton'.}.  Clock constraints on the edges and clock invariants at the states restrict the possible progressions of the system. 
Multiple clocks are needed to allow for the scenario that a drug affects the transition to possible next states in different ways. 
Possible runs and therapies of a timed CHA now include the clock values.  
An extension of CTL, Timed CTL, is used to specify extended goals about the system. 

 In \cref{sec: therapy}, we discuss the problem of automatically generating therapies, i.e., controller synthesis for \CHAs. 
 For simple untimed CHAs this is a well-studied problem and algorithms exist. 
 For timed \CHAs, we show that if we allow only for control at discrete moments in time the problem is decidable for CTL goals. 

Finally, \cref{sec:concl-future-work} concludes with a discussion of several possible extensions of our model, which will be addressed in the future work.

\section{Cancer Hybrid Automata}
\label{sec:cancer-hallmark-automata}
\begin{figure}
\centering
\includegraphics[width=11cm]{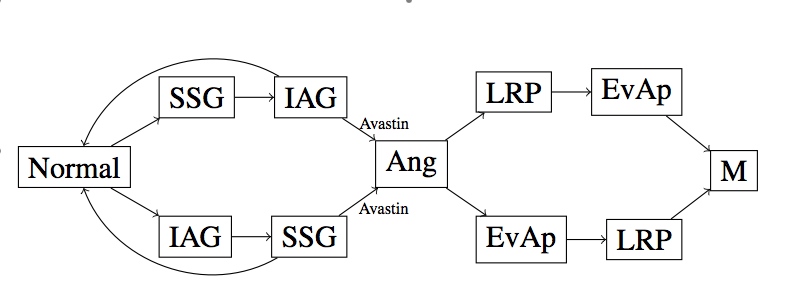}
\caption{A simple CHA whose progression can be stalled by a VEGF-inhibitor such as Avastin.}
\label{fig:example-cha}
\end{figure}

A simple, intuitive example CHA is shown in \cref{fig:example-cha}.
It comprises the following hallmarks (see \cite{hanahan_hallmarks_2000} for more details):
\begin{description}
\item[SSG:] Self-sufficiency in growth signals.
  Roughly speaking, cells no longer depend on external growth-promoting signals, but grow autonomously.
  Usually, such a state is associated with a gain of function of an oncogene or a loss of function of a tumor suppressor gene.
\item[IAG:] Insensitivity to anti-growth signals.
  Cells with this hallmark continue to grow even in the presence of inhibiting signals.
  Usually, certain cell-cycle checkpoints are no longer properly regulated.
\item[Ang:] Sustained angiogenesis.
  This state enables a cancer cell to signal for the construction of blood vessels.
\item[LRP:] Limitless replicative potential.
  While most normal cells can only divide a certain number of times,
  cells with this hallmark can divide without limits.
  In this state, a cancer cell may upregulate telomerase to restore telomere lengths.
\item[EvAp:] Evading apoptosis.
  Normally, cells have a program for controlled cell-death, which is used to remove damaged or otherwise unwanted cells.
  This program is disabled in this hallmark, which allows cells with highly corrupted DNA to survive -- thus facilitating cancer progression further.  
\item[M:] Metastasis. 
This state enables cancer cells to spread from their original location to other parts of the body.
\end{description}

Various possible progressions through these hallmarks can be seen as transitions in the picture
(note that this is a simplified and incomplete model).
For example, Ang can be acquired after SSG and IAG.
Moreover, as mentioned in \cref{sec:introduction}, if a growing tumor fails to acquire Ang, it may starve;
in this case, a solid tumor is unable to grow further and attain the later hallmarks.
For simplicity, it may be modeled as a transition to the normal state.

In this example, the therapy ``give the drug Avastin whenever a state leading up to Ang is reached'' will prevent the cancer from reaching M.

\subsection{Formal model}
\label{sec:formal-model}

In the following, we start with a preliminary and simple formalization of the notions described above.
We will successively extend the formal model in the later sections.

We assume a global set~$D$ of \dfn{drugs}.

\begin{definition}
  A \dfn{Cancer Hybrid Automaton (CHA)} is a tuple
  \[
  H=(V,E,v_0)\mpunct,
  \]
  where
  \begin{itemize}
  \item $V$ is a set of states,%
    \footnote{Strictly speaking, in the case of hallmarks, a state corresponds to a subset of hallmarks that have been acquired.}
  \item $E\subseteq V\times {2^\Drugs} \times V$ is a set of directed edges labeled with sets of drugs, and
  \item $v_0\in V$ is the initial state.
  \end{itemize}
  We usually omit $v_0$ and write just $(V, E)$.
  \end{definition}
Intuitively, an edge $(v,\drugs, v')$ represents a transition
from state~$v$ to state~$v'$ that can be inhibited by any drug from the set~$\drugs\subseteq\Drugs$.
We allow several drugs to be given simultaneously
and refer to such sets $\cocktail\subseteq\Drugs$ of drugs as \dfn{cocktails}.
Given a cocktail \cocktail, the edge $(v, \drugs,  v') \in E$ is \dfnless{inhibited} by~\cocktail if 
$ \cocktail \cap \drugs \neq \emptyset$.
Given a state~$v$ and a cocktail~\cocktail, $v$ can transition to $v'$ under~\cocktail,
in symbols $v \xrightarrow{\cocktail} v'$,
if there is an edge $(v, \drugs, v' )$ that is not inhibited by~\cocktail.
Note that we allow multiple edges (with different labels) between the same two states.
To prevent a transition between two states, all edges connecting them need to be inhibited, 
which is why we need to consider cocktails rather than just single drugs. 
We assume that for every state~$v$ and every cocktail~\cocktail there exists some state~$v'$ such that $v\xrightarrow{\cocktail}v'$
(possibly $v' = v$, these edges were omitted in \cref{fig:example-cha}).

A \dfn{run} of a CHA $H=(V,E,v_0)$ is a sequence of transitions in $E$.
Let $\Runs(v, H)$ denote the set of runs that start in $v$.
We write $\Runs(H)$ for $\Runs(v_0, H)$, 
and by $\Runsf(v,H)$ we denote the set of finite runs from  $\Runs(v,H)$.  

We now formalize how it is possible to \emph{interfere} with the progression of the system.
\begin{definition}
  \label{dfn:therapy}
  A \dfn{therapy} is a function $\thera: \Runsf(H) \rightarrow 2^\Drugs$.
  A \dfn{possible execution} of $\thera$ in $H$ is a run
  \[
  \exec = v_0 v_1 v_2\dots \mpunct,
  \]
  such that for each $i\geq 0$, $v_i\xrightarrow{\thera(\exec_i)}v_{i+1}$,
  where $\exec_i$ denotes the initial segment of $\exec$ up to step $i$. 
\end{definition}

\begin{definition}
  \label{dfn:costs}
  \dfn{Costs} are given by the following (overloaded) function, for some finite dimension~$n$:
  \begin{itemize}
  \item $c:V\to\REAL_{\geq0}^n$ specifying costs of states,
  \item $c:2^\Drugs\to\REAL_{\geq0}^n$ specifying costs of cocktails. 
  \end{itemize}
  
  Thus, both states and cocktails have costs assigned to them,
  represented as $n$-dimensional vectors.
  Dimensions may include toxicity of the drugs, monetary cost of the drugs, discomfort for the patient, etc. 
  
  The cost of a possible execution 
  $
  \exec = v_0 v_1 v_2 \dots
  $
  of a therapy $\thera$ with \dfnless{discount factor}~$0<\delta\leq1$ is
    \[
  c(\exec, \thera, H)=\sum_{i\geq0}\delta^i\big(c(v_i)+c( \thera(\exec_i))\big)\mpunct.
  \]
  The set of possible costs of $\thera$ for a CHA $H$ is
  \[ 
  c(\thera, H) = \{c(\exec, \thera, H)\mid \exec \text{ is possible execution of }\thera \text{ in }H\}.  
  \]
\end{definition}

Now that we have a definition of the set of possible costs of a therapy,
we can compare different therapies with respect to their costs. 

\begin{definition}
  A cost vector $x\in\REAL^n$ \dfn{Pareto-dominates} another vector $x'\in\REAL^n$,
  in symbols $x\prec x'$,
  iff for each $1\leq\ell\leq n$ we have $x_\ell\leq x'_\ell$
  and for some $1\leq\ell\leq n$ we have $x_\ell< x'_\ell$.

  A therapy~$\thera$ \dfn{Pareto-dominates} a therapy~$\thera'$ in a CHA $H$ if
  for each $x\in c(\thera, H)$ and $x'\in c(\thera', H)$
  we have $x\prec x'$.
  The set of \dfn{candidate therapies} for $H$ is
  \[
  \candthera(H)=\{\thera\suchthat \text{$\thera$ is not Pareto-dominated in $H$}\}\mpunct.
  \]
\end{definition}

  For the special case of $1$-dimensional costs (or if there is a function to aggregate cost vectors into single numbers),
  the set of candidate therapies is the set of therapies whose best-case
  cost is not higher than some other therapy's worst-case cost.

This definition of a set of candidate therapies is a very conservative one,
in that it includes any therapy that is not overtly worse than some other therapy.
There are different possibilities for defining the set of candidate therapies,
or for pruning the set further. Examples of such strategies for pruning the set further include \emph{maximin},
i.e., choosing those strategies that lead to the best worst-case outcome,
or \emph{maximax},
i.e., choosing those strategies that lead to the best best-case outcome.
However, making these decisions depends on the risk attitude of patient and doctor
which may not be fully formalizable. Therefore we include all the potentially relevant therapies in the set of candidate therapies. 

\medskip

In order to be clinically applicable, a CHA model may need to be \emph{personalized} for any given patient or cancer type.
This personalization will result in families of \CHAs,  with different sets of candidate therapies.
While we will not give full details here, we wish to describe one possible application for such richer models.

For families of automata, we can ask whether there are any \emph{universal} therapies
for all of the included automata. Such therapies can result in faster and cheaper treatments.

To be able to apply therapies across different automata, their domain must be the same. 
This requirement can be satisfied, for example, by considering \CHAs that contain the same set of hallmarks, 
and therapies that either depend only on the current state, or that have the set of all sequences of states as domain. 
The following definition applies to therapies on such unified domains. 

\begin{definition}
  Given a family~\HH of \CHAs, the set of \dfn{(universal) candidate therapies} for~\HH is
  \[
  \candthera(\HH)=\bigcap_{H\in\HH} \candthera(H)\mpunct.
  \]

  A set~$\TT$ of therapies \dfn{covers}~\HH if
  \[
  \TT\cap\candthera(H)\neq\emptyset\text{ for all $H\in\HH$}\mpunct.
  \]
\end{definition}
Note that if $\candthera(\HH)\neq\emptyset$ then for each $\thera\in\candthera(\HH)$, $\{\thera\}$ covers \HH.

\subsection{Temporally extended goals: CTL}
We have seen in the previous section that therapies can be compared according to their costs.
Thus, the problem of finding the right therapy can be viewed as an optimization problem.
It can, however, be necessary to have more detailed control over the therapeutic objectives.
Simple reachability properties can be used as goals, such as ``metastasis must never be reached''. 
For more expressivity we can use Computation Tree Logic (CTL)~\cite{springerlink:10.1007/BFb0025774} to specify goals.

\begin{example} The goal
$\AG \neg \text{M}$
states that metastasis is never reached.
Another possible goal could be 
\[\AG(\text{Ang} \rightarrow \AG\neg \text{EvAp})\mpunct. \]
This sentence means that whenever sustained angiogenesis is acquired,
then at no point in the future the capability of evading apoptosis will be obtained.
\end{example}

One may be interested in checking properties of the CHA itself, without application of a therapy.
This goal can be achieved by using CTL model checking  (see, e.g.,~\cite{clarke_model_1999}).
CTL properties can also be checked on the possible executions of a given pair of therapy and untimed CHA.
Supervisory control for finite automata with CTL goals is known to be EXPTIME-complete, and 
controller synthesis algorithms exist \cite{jiang_supervisory_2006}.


\bigskip

The above representation of a cancer automaton is intuitive, but it does not include \emph{timing}.
It fails to model the fact that
some transitions could be very short while others may take many years.
In the next section we introduce timed \CHAs, 
which are automata equipped with a set of real-valued variables, denoted as \emph{clocks}, and constraints on the edges and states restrict the progression of the system. 
This model will be a special kind of hybrid automaton, justifying the word \emph{hybrid} in `cancer hybrid automata'.

\section{Timed \CHAs}
\label{sec:timed-cha}
The framework we built so far is somewhat idealized in that transitions occur spontaneously and drugs can switch off transitions completely.
More realistically, transitions would take certain durations of time, and drugs can slow down (or stop) the transition process.
For example, in pancreatic cancer,
it takes about a year for K-\textit{ras} mutations in a cell to lead to neoplasms (so-called PanINs)~\cite{hruban_progression_2000}.
To model durations, we will now add a notion of \emph{time} to our CHA framework.

We start with the assumption that the acquisition of a hallmark requires a certain minimum amount of time.
We do not specify exactly how that time is determined,
but it could be the stopping time of a stochastic process such as randomizing over a set of driver mutations,
or some value obtained from clinical data.
Only after that time a given transition will be possible, and as mentioned, drugs can be used to prolong this time.

Further, we allow states to have \emph{invariants},
specifying the maximum time that the system can remain in the respective state.
For example, a tumor may only be able to remain in a state of unbounded growth without angiogenesis for a certain number of months.

\begin{figure}
\centering
\includegraphics[width=11cm]{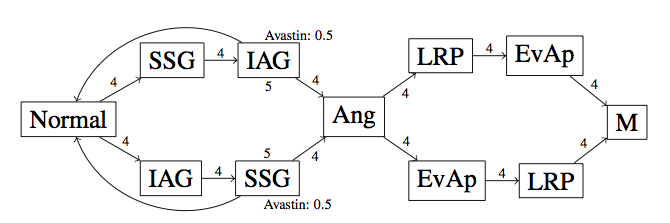}
\caption{A simple timed CHA.
  The edges are labeled with the minimum times needed to make the respective transitions.
  In the two states that lead up to Angiogenesis, Avastin can be given to slow down the progress by half.
  Those states are labeled with invariants, and depending on the precise timing,
  these invariants can force the system back to Normal before the transition to Angiogenesis is possible.}
\label{fig:example-timed-cha}
\end{figure}

\Cref{fig:example-timed-cha} shows the automaton from \cref{fig:example-cha} with timing information added,
illustrating this intuition.
We formalize the extension in the following.

We assume a finite set $X$ of real-valued variables called \dfnless{clocks}, over which the set of \dfnless{constraints} $\mathcal{C}(X)$ is generated according to the grammar
\[
\phi ::= x \geq k \mid \phi \wedge \phi\mpunct,
\]
where $k \in \NAT$ and $x \in X$.
A \dfnless{valuation} of the variables in $X$ is a mapping $\val: X \to \REAL_{\geq 0}$.
We denote the null valuation $x\mapsto 0$ by $0$.
By $\val \models \phi$ we denote that $\val$ satisfies $\phi$.

\begin{definition}
  \label{dfn:timed-cha}
  A \dfn{timed CHA} is a tuple $H = (V, E, v_0, \Inv, \rho)$ where
  \begin{itemize}
  \item $V$ is a set of states, 
  \item $E\subseteq V\times \mathcal{C}(X) \times V$ is a set of directed edges each labeled with a clock constraint,
  \item $v_0\in V$ is the initial state, 
  \item$\Inv: V \times X \rightarrow \NAT$ is a partial function specifying the time limit (if any) for each clock that the system can remain in a given state (this is also called the \emph{invariant}), and
  \item $\rho: V \times \Drugs \times X \to \REAL_{\geq 0}$ yields a function specifying how a given drug influences the clocks at a given state.  
  \end{itemize}
\end{definition}

Intuitively, at a given state $v$, the drug $\drug$ modifies the clock rate, by slowing down or speeding up the clock $x$ as specified by a multiplicative factor $\rho(v, \drug, x)$.
When the factor is $1$, the drug has no effect on that clock, and when it is $0$, it effectively stops the clock from progressing. 
If several drugs have an effect on a clock, their factors are multiplied.
We extend~$\rho$ to cocktails by setting $\rho(v,\cocktail,x)=\prod_{\drug\in\cocktail}\rho(v,\drug,x)$ for any cocktail $\cocktail \neq \emptyset$, and by convention, $\rho(w,\emptyset,x) = 1$.

A directed edge $(v, \phi, v')$ represents a transition from $v$ to $v'$
that can take place once the time constraint $\phi$ is satisfied.

We assume that for each state $v$ that has a time limit for a clock $x$,
there is an outgoing edge $(v, \phi, v')$ such that $\val \models \phi$ for all $\val$ with $\val(x) = \Inv(v, x)$.%
\footnote{Note that this requires $\val\models\phi$ even for valuations that exceed some other clock's invariant;
  however, this does not have an effect since we only allow $\geq$ constraints on the edges.}
This edge specifies the behavior of the system if the respective clock reaches its time limit.

\medskip

The cost functions in the context of timed \CHAs are the same as those for the untimed version,
but with a timed interpretation:
$c(v)$ is the cost of staying at state $v$ per time unit (days/weeks/months/years),
and $c(\cocktail)$  is the cost of administering a drug cocktail $\cocktail$ per time unit.

\medskip

We next see how to adapt the definitions related to runs of a CHA to the timed version,
starting with the notion of a \emph{timed state}.

\begin{definition}
  \label{dfn:timed-cha-evolution}
  A \dfn{timed state} of a timed CHA $(V, E)$ is a tuple $(v, \val)\in V\times\REAL^X$, where
  $v$ is a state and $\val$ a clock valuation.
  There are two types of \dfn{transitions} between timed states:
  \begin{enumerate}
  \item \dfnless{Delay} transitions, in symbols $( v, \val) \xrightarrow{\delta,\cocktail}(v, \val')$, where
    \begin{itemize}
    \item $\delta\in\REAL_{>0}$ represents the (real) time delay,
    \item \cocktail denotes the cocktail active during that time,
    \item $\val'(x)=\val(x)+\delta\rho(v,\cocktail,x)$ for all~$x$, and
    \item $\val'(x)\leq\Inv(v,x)$ for all $x$ with $\Inv(v,x)$ defined.
    \end{itemize}
  \item \dfnless{State} transitions, in symbols $( v, \val) \rightarrow (v', 0)$, where
    \begin{itemize}
    \item there is an edge $(v, \phi, v') \in E$  with $\val\models\phi$.
    \end{itemize}
  \end{enumerate}
\end{definition}
Note that whenever a state transition takes place, the clocks are reset.
This strategy simplifies our presentation and could be replaced by explicit clock resets as common in the literature.

This setup includes the special case where there is one clock unaffected by any drug, representing real time.
Invariants over that clock can be used to specify, for example, the duration over which the tumor can remain in a certain state. 

This timed setup can also emulate the concept of edges labeled with drugs that inhibit them.
This model can be constructed as follows: 
Suppose we want to model an edge between two states $v, v'$ that can be inhibited by a drug $\drug$. 
Then we can introduce a clock variable $x_{\drug, v'}$ with $\rho(v, \drug, x_{\drug, v'}) = 0$,
and add a constraint $x_{\drug, v'}\geq z$ to the edge between~$v$ and~$v'$, for some~$z>0$.
As long as drug~$\drug$ is given before the constraint is satisfied, the transition will be inhibited.
However, once the constraint is satisfied, the tumor has advanced too far and it is no longer possible to inhibit the transition.

\medskip

A \emph{run} in the case of a timed CHA $H$ is a non-Zeno%
\footnote{That is, not containing an infinite chain of timed transitions with convergent total duration.}
sequence of delay and state transitions.
Similar as before, let $\Runs((v,\val), H)$ denote the set of runs that start in $(v, \val)$.
We write $\Runs(H)$ for the set $\Runs((v_0, 0), H)$,
and $\Runsf((v, \val), H)$ for the set of finite runs from $\Runs((v,\val), H)$.  
 
 \begin{definition}
A \dfn{therapy} is a function $\thera: \Runsf(H) \rightarrow 2^\Drugs$.  A \dfn{possible execution} of $\thera$ in $H$ is a run
\[ \exec = (v_0, 0) (v_1, \val_1) (v_2, \val_2) \cdots
\]
such that for all $i$ with delay transitions $(v_i,\val_i)\xrightarrow{\delta,\cocktail}(v_{i+1},\val_{i+1})$,%
\footnote{Note that $v_i=v_{i+1}$.}
for every $0\leq \delta' < \delta$
\[ \thera( (v_0, 0)\ldots (v_i, \val_i)(v_i, \val_i+\delta'\rho(v_i,C))) = \cocktail,
\]
where $\rho(v_i,\cocktail)$ denotes the partial evaluation of~$\rho$, i.e., the function~$x\mapsto\rho(v_i,\cocktail,x)$.

\end{definition}
This last condition ensures that the therapy does not change during a transition,
or, put differently, that a change in therapy is always reflected by starting a new transition.

For any finite run $\run\in\Runsf(H)$, we denote its \dfnless{duration} as
\[
\tau(\run)=\sum_{0\leq j<len(\run)}
\begin{cases}
  \delta & \text{if $\run_j\xrightarrow{\delta,\cocktail} \run_{j+1}$ for some $\delta,\cocktail$}\\
  0 & \text{otherwise,}
\end{cases}
\]
where $len(\run)$ denotes the length of the state sequence in~$\run$ and $\run_i$ its initial segment of length~$i$.

\begin{definition}
\label{dfn:costs-timed-cha}
  Given a CHA $H$ and a possible execution $\exec$ of a therapy $\thera$,
  the \dfn{cost} of $\exec$ given $\thera$ with discount factor~$0<d\leq1$ is 
 \[
  c(\exec, \thera, H) =   \sum_{i\geq0} \frac 1d\left(e^{-d\tau(\exec_i)}-e^{-d\tau(\exec_{i+1})}\right)(c(v_i)+c(\thera(\exec_i)))
  \]
  (as before, by $\exec_i$ we denote the initial segment of $\exec$ up to step $i$).
  This simple discounting function does not necessarily capture a real patient's preferences,
  but any convergent function will work in its stead.
  We will consider more realistic functions in the future,
  which can potentially be designed on a case-by-case basis depending on the patient's valuation.

  The set of possible costs of $\thera$ in a timed CHA $H$ is the set of costs of possible executions of $\thera$,
  \[ 
  c(\thera, H) = \{c(\exec, \thera, H)\mid \exec \text{ is possible execution of }\thera \text{ in }H\}. 
  \]
\end{definition}

The notions of Pareto dominance and universal therapies carry over from untimed \CHAs.

\subsection{Timed CTL}
We can extend the CTL goals of the previous section to include time~\cite{Alur19932}.
For example, the goal $\AG_{\leq 20} \neg \text{M}$
says that metastasis is not reached within $20$ time units (e.g., 20 years).
This kind of goal represents the approach of turning cancer into a chronic disease, rather than trying to cure it completely.
For example, the above formula may be appropriate for a patient of sixty years of age, who may then be able to get a less strenuous therapy, 
while for a younger patient the time requirements may be more extensive. 

Out of all the therapies satisfying a CTL goal, the best ones may be chosen either by a separate cost optimization,
or by incorporating cost requirements into the formulas using a weighted version of CTL~\cite{Patricia20063}.

\section{Automatic therapy design for  \CHAs}
\label{sec: therapy}
Given the complexity of (timed) cancer progression and the influence of various drugs,  the task of finding near-optimal therapy plans is (soon to be) beyond manual planning, 
and automated computational tools are very desirable. 

The controller synthesis problem for different classes of automata have been studied in the literature,
often restricted to achieving \emph{safety} (avoiding a set of `bad' states) and \emph{reachability} (eventually reaching a `goal' state) properties.
Such properties form a sub-class of what can be expressed in richer temporal logics such as CTL.
Safety properties are especially relevant for \CHAs, because goals such as ``metastasis will never be reached'' can be expressed.

Untimed CHAs are a special kind of discrete automata for which efficient controller synthesis algorithms exist 
and can be applied to automatically design therapy-plans
(see e.g. \cite{maler_onthesynth} for control using safety goals and  \cite{jiang_supervisory_2006} for an algorithm that uses CTL specifications).

\paragraph{Control of timed \CHAs}
For timed CHAs, however, control is not as straightforward.
CHAs are a special class of hybrid automata. 
Unfortunately, in hybrid systems, even simple verification and control problems like reachability and safety are undecidable \cite{Henzinger:1995}.
However, several decidable subclasses of hybrid automata exist for which algorithms have been devised. 
One such subclass is that of \emph{rectangular} hybrid automata. 
A rectangular automaton is an automaton in which the clock constraint on each edge is a rectangular region of continuous states.
That is, it specifies for each clock a (possibly unbounded) interval that should contain its value. 
Also, the clock speed at each state is assumed to be bounded from below and above. 

Rectangular automata form a most general class of hybrid automata for which even the reachability model checking problem is decidable 
\cite{Henzinger:1995, Henzinger99rectangularhybrid} and controller synthesis algorithms have been developed.
For example, in \cite{Henzinger99rectangularhybrid} Henzinger et al. show that the control problem with LTL specifications is 
EXPTIME-complete in the size of the game,
 and 2EXPTIME-complete in the size of the formula.

These results rely on the requirement that the rectangular hybrid automata satisfies a property called \emph{initialization} or \emph{constant reset}.
Initialization states that whenever the speed of a clock changes after a transition, 
the value of the variable is reinitialized to a fixed value (or a value in a fixed interval). 
This property cannot be relaxed without making the control problem undecidable \cite{Henzinger:1995}: 

\paragraph{From timed \CHAs to rectangular hybrid automata}
Timed CHAs bear a striking resemblance to rectangular hybrid automata,
and it is thus worth exploring whether some of the controller synthesis results and algorithms can be applied to CHA models as well.
Unfortunately, existing decidability results do not carry over directly because of some important differences between CHAs and (rectangular) hybrid automata.

First, in the hybrid automata literature, the rates of the clocks are generally assumed to be constant at any given state%
\footnote{One exception are so-called \emph{differential games}~\cite{maler_control_2002}, but their theory has not been well developed.}
and what is controllable are (some of) the transitions between states.
In the CHA framework, in contrast, the rates of the clocks is what can be affected by control actions (drugs),
while the transitions (tumor progression) cannot be directly manipulated.
However, this difference is mainly conceptual as a timed CHA can be translated to a hybrid automaton as follows:

Given a set of drugs~$\Drugs$ and a CHA $H$ with states~$V$,
we construct a hybrid automaton~$\RH$ in the following way:
For each state $v \in V$ and each cocktail $\cocktail \in 2^\Drugs$, $\RH$~contains a state $v_\cocktail$
with the same clock invariants as~$v$.
For any edge between two states $v, v'\in V$,
$\RH$~contains an uncontrollable edge between $v_\cocktail$ and $v'_\cocktail$, for each cocktail $\cocktail$,
with the same clock constraints and resets as on the CHA edge.
In addition to the uncontrollable edges, there are controllable directed edges from $v_\cocktail$ to $v_{\cocktail'}$
for each $v$, $\cocktail$ and $\cocktail'$. 
These edges represent changes of therapies, and have no clock constraints or resets.
At a state $v_\cocktail$, the rate of each clock $x \in X$ is fixed, given by $\rho(v, \cocktail, x)$.
This translation yields an automaton of size exponential in the number of drugs, but linear in the number of CHA states.

The result is a rectangular hybrid automaton. 
However,  the translated CHA does not satisfy initialization,
 as the clock values (indicating progression time) are kept along controllable (change of cocktail) transitions
while changing the rates of the clock. 
Thus, the negative results of Henzinger et al. \cite{Henzinger99rectangularhybrid} are no longer applicable. 

\paragraph{Discretized control}
The simplest way around the undecidability of the control problem for rectangular hybrid automata that do not satisfy initialization 
is to allow for control moves (in our case, therapeutic interventions) only at discrete instants of time. 
Henzinger and Kopke \cite{Henzinger1999369} give an exponential-time algorithm for discrete-time safety control with CTL goals of rectangular hybrid automata with bounded and non-decreasing variables.
They also show the problem to be EXPTIME-hard and discrete-time verification of rectangular hybrid automata to be solvable in PSPACE.

Even though our definition of timed \CHAs does not require clocks to be bounded,
such a restriction would not impose a severe limitation.
By bounding the clocks by some value that even the healthiest patient will never reach,
we can thus aim for decidability without forfeiting any meaningful therapy.
The algorithms from \cite{Henzinger1999369} do not directly apply to  \CHAs as their framework requires all discrete transitions to be controllable, 
whereas our cancer progression transitions are uncontrollable.
However, they can be extended to include our framework via the following theorem, for which
we only provide a sketch of the proof. The full proof can be found in the extended version of this paper. 

\begin{theorem}[Discrete control of bounded \CHAs]
The controller synthesis problem of bounded discretized \CHAs for CTL formulae can be solved in EXPTIME.
\end{theorem}

\begin{proofsketch}
First, we can translate the bounded CHA $H$ into a rectangular hybrid automaton $\RH$ as described earlier. 
Then, the rectangular hybrid automaton $\RH$ can be described as a hybrid \emph{game}
\footnote{A game automaton is an automaton in which two players can make discrete moves. 
In our case not only the controller but also nature/the cancer can make discrete moves.}
$\HG$ by specifying that the controller is only allowed to make moves that include a change of therapy:
from $c$ to $c'$ at state $v$ by moving from $(v,c)$ to $(v,c')$, 
and cancer is only allowed to pick an accessible new CHA state from the available $((v,c)(v',c))$ transitions.

Next we can extend the discretization method as given in \cite{Henzinger1999369} for rectangular automata to hybrid games. 
We can define a \emph{sampling control game} \DHG in which the players can only make one move every time unit, 
by adding a new variable $x_{n+1}$ such that $\Inv(v, x_{n+1}) = 1$ at each state; each clock constraint $\phi$ in the automaton becomes $\phi \wedge x_{n+1} \geq 1$ ($0 \leq x_{n+1} \leq 1$); 
the rate of $x_{n+1} $ is $1$ at all states; and $x_{n+1}$ is reset to $0$ after each discrete transition.
This construction guarantees that moves by the cancer and therapist are always followed by a delay transition of duration $1$
\footnote{Note, you have to assume that the automton is big enough: in the original automaton it is not possible to make two moves in one time unit. }. 

We can then define a bisimulation relation on the states of the discretized hybrid game $\DHG$ as in \cite{Henzinger1999369} as follows:
We define an equivalence relation $\approx_n$ on $\REAL^n$ (the set of clock valuations) such that 
$\mathbf{y} \approx \mathbf{z}$ iff $\lfloor y_i \rfloor = \lfloor z_i \rfloor$ and $\lceil y_i \rceil = \lceil z_i \rceil$ for all $a \leq i \leq n$. 
Now, given two states $(v, \val)$ and $(v', \val')$ we define  $(v, \val) \cong_{\DHG} (v', \val')$ if $v = v'$ and $\val \approx \val'$.
 (we can also define $(v, \val) \cong_{\DHG}^m (v', \val')$ for a bound $m$ ).
We can show that this is indeed a bisimulation preserving CTL satisfaction, 
and since the result is a finite representation (exponential in size) of the original CHA, 
it follows that control of discretized bounded \CHAs with CTL goals is solvable in EXPTIME.
\end{proofsketch}

\section{Conclusions}
\label{sec:concl-future-work}

This paper establishes a general formalism for describing cancer progression,
without relying on any detailed mechanistic model of cancer pathways (which can be included independently as models of the discrete states).
Our goal was to design a conceptually clear framework based on realistic biological foundations.
As a case study, we have used this model to describe cancer hallmarks and their dynamics. 

We discuss below how our framework can be used, as is, to model phenomena beyond what we discussed so far. 
Then, we point out the limitations of the current paper and give a list of topics that we plan to address in the near future.

\subsection{Modeling growth, heterogeneity and anti-hallmarks}
\label{sec:model-heter}

\paragraph{More general clocks:}
Thus far, we have referred to the clocks in \CHAs as measuring \emph{time}. 
However, they could be measuring different properties like \emph{tumor size}, 
\emph{motility} or \emph{spatial properties}. 
For example, in the case of tumor size, the growth rate of the tumor may depend on the current discrete states of the progression and 
drugs can influence this rate. With this model we can reproduce the tumor growth dynamics as described in \cite{Rodriguez}, by introducing two clocks: 
one measuring the number of stem cells and the other the number of differentiated cells. 
The various mutations can be modeled as transitions to a next state with different growth dynamics depending on the mutations already acquired.

\paragraph{Heterogeneity in tumors:}
So far we have modeled states of a CHA as representing the unique dominant phenotype of the tumor cell population.
However, most forms of cancer are not likely to be monoclonal,
i.e., consist of only one population in which the clonal expansions postulated by Hanahan and Weinberg take place,
but rather involve several sub-populations of tumor cells~\cite{navin_tumour_2011}, each with a distinct dominant phenotype~\cite{fidler_tumor_1978,heppner_tumor_1984}.
In order to model this heterogeneity, we can simply think of a CHA state as representing a \emph{vector} of dominant phenotypes,
one for each sub-population.
One or several components of such a vector may differ from one state to the next,
corresponding to a change of the dominant phenotype in the corresponding sub-population(s) during the respective transition;
or the length of the vector may change,
corresponding to new distinct sub-populations emerging or existing sub-populations dying out.
This approach is, however, rather crude in modeling tumor heterogeneity,
and does not straightforwardly accommodate, for example,
information about tumor geometry or a model of the resulting spatial effects.

\begin{figure}
\centering
\includegraphics[width=11cm]{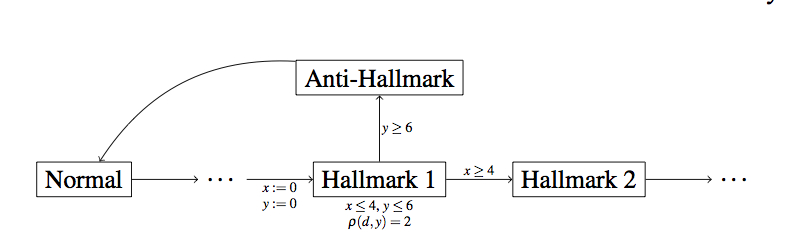}
\caption{Illustrating how to model an anti-hallmark using two clocks~$x$ and~$y$
    and a drug~$d$ that speeds up clock~$y$ at Hallmark~$1$ by a factor of~$2$.}
\label{fig:example-anti-hallmark}
\end{figure}


\paragraph{Anti-hallmarks}
Instead of trying to \emph{slow down} cancer progression,
there has recently been growing interest in approaches to \emph{speed up} the process to a degree
which will make the tumor nonviable and ``push it over the edge'' towards collapse.
We refer to such nonviable states as \emph{anti-hallmarks}.
They can be modeled by putting constraints on the transitions leading to them
that will never be satisfied, unless a drug is given which speeds up a certain clock.
For example, consider the CHA in \cref{fig:example-anti-hallmark}.
At Hallmark~$1$, without interference (both clocks increase with rate~$1$),
the transition to Hallmark~$2$ will be taken after~$4$ time units.
A drug that speeds up clock~$y$ by a factor of~$2$ will instead push the tumor to the Anti-Hallmark state,
if given starting at most $1$~time unit after entering Hallmark~$1$.

\subsection{Extensions and Future Work}
\label{sec:future-work}

\paragraph{Partial observability and tests:}
The framework introduced in this paper assumes perfect information about the state of the system.
In reality however, a clinician will only have partial observations of the tumor's internal state.
To reduce uncertainty about the current state of the cancer progression, \emph{tests} can be performed.
Our formal framework can be extended to include partial observability and tests,
both for untimed and timed \CHAs.
Partial knowledge about the tumor's internal state can be modeled by introducing the notion of a belief set. 
Tests can be incorporated into the definition of a therapy as actions that reduce uncertainty about the current state. 
A therapy can then be described as a function from the set of belief-runs to cocktails or tests. The details appear in the full paper.

\paragraph{Compositional models:}
In a patient, cancer itself is not the only system of relevance.
Other systems interact with the tumor's development, and especially during a therapeutic intervention, they need to be monitored.
For example, the immune system and its role throughout carcinogenesis are receiving more and more attention~\cite{citeulike:457440},
and the liver needs to be monitored to avoid damage due to excess toxicity.
In principle, other subsystems of an organism could be modeled as hybrid automata in the same way as our CHA,
which could then be composed to an overall model for which therapies with goals spanning all subsystems could be generated.
\paragraph{}
Building on our conceptual foundation, we plan to address several important issues next.

\paragraph{Algorithmic issues:}

In section \ref{sec: therapy}, we have shown that the controller synthesis problem for timed CHAs is decidable
 if both the therapist and the cancer are only allowed to make moves at discrete moments in time. 
In the future, we plan to focus more on to the algorithmic side of
verifying cancer hallmark automata, automatically generating therapies (including cost minimization), finding promising drug targets, etc.

\paragraph{Model extraction:}
Finally, we omitted a description of the methodologies needed for extracting cancer phenotypes
and their temporal progression models from data or mechanistic pathway and population models.
For example, there is currently no consensus that the cancer hallmarks described in the literature constitute a complete list,
nor is there a clear understanding (either phenomenologically or mechanistically) of their precise discrete dynamics.
We also believe that spatial structure (geometry, growth curve, spatial distribution of heterogeneity, etc.)
as well as motility (self-seeding, circulating tumor cells) may hold additional and important clues
that can be easily incorporated into our therapy design~\cite{comen_clinical_2011,norton_cancer_2008}.
Therefore, we plan to extract models from spatio-temporal data, for example,
data obtained from detailed simulations, or gene expression and imaging data from patients or mouse models. 
We plan to use statistical inference algorithms for model extraction (such as GOALIE~\cite{ramakrishnan_reverse_2010})
in order to reconstruct temporal (or spatio-temporal) phenomenological models of cancer-related processes from such data.

\nocite{*}
\bibliographystyle{eptcs}
\bibliography{HSB2012}

\end{document}